\documentclass[amsmath,aps,prc,twocolumn,groupedaddress,showpacs]{revtex4}
\usepackage{graphicx}

\begin{document}

\title{Impact of nuclear ``pasta'' on neutrino transport in
collapsing stellar cores}

\author{Hidetaka Sonoda$^{a,b}$, Gentaro Watanabe$^{c,b}$,
 Katsuhiko Sato$^{a,d}$, Tomoya Takiwaki$^{a}$,
 Kenji Yasuoka$^{e}$, and Toshikazu Ebisuzaki$^{b}$}
\affiliation{
$^{a}$Department of Physics, University of Tokyo, Tokyo 113-0033,
Japan
\\
$^{b}$The Institute of Chemical and Physical Research (RIKEN),
Saitama 351-0198, Japan
\\
$^{c}$NORDITA, Blegdamsvej 17, DK-2100 Copenhagen \O, Denmark
\\
$^{d}$Research Center for the Early Universe, University of Tokyo,
Tokyo 113-0033, Japan
\\
$^{e}$Department of Mechanical Engineering, Keio University, Yokohama,
223-8522, Japan
}

\date{\today}

\begin{abstract}
Nuclear ``pasta'', nonspherical nuclei in dense matter, is predicted to 
occur in collapsing supernova cores. We show how pasta phases
affect the neutrino transport cross section via weak neutral current 
using several nuclear models.
This is the first calculation of the
neutrino opacity of the phases with rod-like and slab-like nuclei 
taking account of finite temperature effects, 
which are well described by the quantum molecular dynamics.
We also show that pasta phases can occupy 10--$20\ \%$
of the mass of supernova cores in the later stage of the collapse.
\end{abstract}


\maketitle

Interactions between neutrinos and matter 
are one of the key elements for understanding the mechanism of 
core-collapse supernovae. 
In the collapse phase, neutrino-matter interactions play a crucial
role for explosion: the efficiency of the trapping of neutrinos inside 
the collapsing core governs its lepton fraction $Y_{L}$ (the ratio of
the number density of leptons to that of baryons) and hence 
the mass of the inner core at bounce. The latter quantity determines the 
initial position and the strength of the shock wave \cite{bethe_rev}.

In the neutrino-trapping stage, 
interactions between neutrinos and matter are dominated by coherent
scattering of neutrinos from heavy nuclei via the weak neutral current, 
which we focus on in this article.
In the case of scattering of neutrinos by an isolated nucleus,
neutrinos are coherently scattered by nucleons in the nucleus
provided the wavelength of neutrinos is much larger than the radius of
the nucleus $r_{N}$.
Therefore, roughly speaking, the cross section per nucleon is enhanced
by a factor of $A$ \cite{freed74,sato}, 
where $A$ is the mass number of the nucleus, 
due to the fact that the scattering is coherent.
However, when neutrinos are scattered by matter in the core,
not by an isolated nucleus, more than one nucleus would contribute 
to the scattering amplitude and the coherence can be disturbed 
if the internuclear distance is comparable to or smaller than 
the wavelength of neutrinos \cite{Itoh75}.
This correction is due to the spatial correlation among nuclei. 
Especially, at higher densities, nuclei form a lattice because of the
strong Coulomb forces among nuclei and thus the correlation among them
is very different from that of the liquid phase at lower densities.

More interestingly, 
the nuclear shape itself would change from sphere to rod or slab, etc.,
at subnuclear densities in the collapsing supernova core 
\cite{rpw,hashimoto,williams,lassaut,qmd_hot,qmd_transition}.
These nonspherical nuclei are referred to as ``nuclear pasta''.
Transitions to pasta phases also change the correlation among nuclei
and hence the neutrino opacity.

At the present time, the effects of nuclear pasta on the neutrino opacity
have yet to be understood completely. 
One of the important questions to be clarified is how the neutrino
opacity for the conventional models of dense matter 
used in supernova simulations
is modified by taking account of the pasta phases.
Horowitz {\it et al.} have studied a related problem using the
framework of quantum molecular dynamics (QMD) \cite{horowitz}, but
they have not reproduced the phases with rod-like and slab-like nuclei
and those with rod-like and spherical bubbles probably due to the use of
a short screening length ($10$ fm) compared to the internuclear distance
(see a discussion in Ref.\ \cite{tours}).
Therefore the above question is still open.

We calculate the cross section for elastic neutrino scattering 
from pasta phases via weak neutral current \cite{note_elastic}.
After averaging over the spin states of nucleons,
resulting expression of the neutrino cross section per nucleon 
to the first order is
\begin{eqnarray}
 \frac{1}{N}\frac{d\sigma}{d\Omega}(\mathbf{q})
 =\frac{G_F^2E_{\nu}^2}{4\pi^2}
 (1+\cos\theta){c_v^{(n)}}^2\overline{S}(\mathbf{q}),\label{cs of coh}
\end{eqnarray}
where $N$ is the number of nucleons in the system,
$G_F$ is the Fermi coupling constant,
$E_{\nu}$ is the energy of neutrinos, $\theta$ is the scattering angle,
${c_v^{(n)}}$ $({c_v^{(p)}})$ is the vector
coupling constant of neutrons (protons) to the weak neutral current
(${c_v^{(n)}}=-1/2,\ {c_v^{(p)}}=1/2-2\sin^2 \theta_W,\ \sin^2\theta_W=0.23$),
$\hbar\mathbf{q}$ is the momentum transfer, and $\overline{S}(\mathbf{q})$ is
defined as follows:
\begin{widetext}
\begin{eqnarray}
 \overline{S}(\mathbf{q})=\frac{1}{{c_v^{(n)}}^2}\left\{
 {c_v^{(n)}}^2(x_n^2 S_{nn}(\mathbf{q})+x_p x_n)+{c_v^{(p)}}^2
 (x_p^2 S_{pp}(\mathbf{q})+x_p x_n)+{c_v^{(n)}}{c_v^{(p)}}x_p x_n
 (S_{pn}(\mathbf{q})+S_{np}(\mathbf{q})-2)
 \right\}.\label{def S}
\end{eqnarray}
\end{widetext}
In Eq.~(\ref{def S}), $x_p$ $(x_n)$ is the proton (neutron) fraction
of the system, $S_{AB}(\mathbf{q})$ with $A, B=n$ or $p$ is the partial
static structure factor of nucleons of species $A$ and $B$,
\begin{eqnarray}
 S_{AB}(\mathbf{q})=1+\rho\int d^3r\ e^{i\mathbf{q}\cdot\mathbf{r}}
 \{ g_{AB}(\mathbf{r})-1\},\label{def part S}
\end{eqnarray}
where $\rho$ is the average number density of nucleons in the system 
and $g_{AB}$ is
the two-body distribution function of species $A$ and $B$ defined as follows,
\begin{align}
 &g_{AB}(\mathbf{r})\nonumber\\
 \equiv& \frac{1}{\rho_A\rho_B}\frac{1}{V}
 \int d^3r' \left\langle\psi^{\dag}_A(\mathbf{r}+\mathbf{r}')
 \psi^{\dag}_B(\mathbf{r})\psi_B(\mathbf{r})
 \psi_A(\mathbf{r}+\mathbf{r}')\right\rangle.
 \label{def g}
\end{align}
In Eq.\ (\ref{def g}), $\langle \cdots \rangle$ 
denotes the statistical average, 
$\psi_A$ ($\psi_B$) is the field operator of
particle $A$ ($B$), $\rho_A$ ($\rho_B$) is number density of $A$ ($B$),
and $V$ is the volume of the system.
If we ignore the contribution of the axial vector current,
the cross section for the scattering of neutrinos by a single neutron
is given by
\begin{eqnarray}
 \frac{d\sigma_n}{d\Omega}=\frac{G_F^2E_{\nu}^2}{4\pi^2}
  {c_v^{(n)}}^2(1+\cos\theta).
 \label{cs of one}
\end{eqnarray}
Comparing Eqs.\ (\ref{cs of coh}) and (\ref{cs of one}), we see that
$\overline{S}(\mathbf{q})$ is the amplification factor of
the cross section per nucleon due to the structure of nuclear matter.
We normalize Eq.\ (\ref{def S}) by $c_v^{(n)}$ because neutrons
dominantly contribute to the vector current.

In the field of simulation of supernovae, researchers often use
transport cross section defined as
\begin{eqnarray}
\frac{d\sigma_t}{d\Omega}(\mathbf{q})\equiv\frac{d\sigma}{d\Omega}(1-\cos
\theta),\label{transport_def}
\end{eqnarray}
which means the efficiency of the momentum transport from neutrinos to matter.
Note that the cross section depends on the relative direction between
the momentum of neutrinos and the symmetry axis of the pasta phases 
as shown later in Eq.\ (\ref{Sq_ld}).
However, collapsing cores are polycrystalline; 
the symmetry axes of grains are randomly oriented and 
the size of the grains would be much smaller than the mean free path of neutrinos.
Thus the neutrino opacity of supernova cores would be well characterized by
the angle-average of $\overline{S}(\mathbf{q})$ (an average over the 
scattering angle $\theta$),
\begin{eqnarray}
 \langle \overline{S}(E_{\nu}) \rangle\equiv \frac{3}{32\pi^2}
 \int d\Omega_i d\Omega_f (1-\cos^2\theta)\overline{S}(\mathbf{q}) ,
\label{ave transport}
\end{eqnarray}
where $\Omega_i$ and $\Omega_f$ are integrated over initial and final
directions of neutrinos.
The above quantity
is related to the transport cross section $\sigma_t$ per nucleon as
\begin{eqnarray}
 \sigma_t=\langle \overline{S}(E_{\nu})\rangle \sigma_t^0,
 \label{transport cs}
\end{eqnarray}
where $\sigma_t^0$ is the total transport cross section of one neutron
via the vector current,
\begin{eqnarray}
 \sigma_t^0=\frac{2G_F^2E_{\nu}^2}{3\pi}{c_v^{(n)}}^2.
 \label{transport cs neu}
\end{eqnarray}
Thus $\langle \overline{S}(E_{\nu})\rangle$ is an amplification
factor of the total transport cross section.

When nuclei form a crystalline lattice,
the partial static structure factor $S_{AB}(\mathbf{q})$,
within the Wigner-Seitz (WS) approximation, can be written as
\begin{align}
 S_{AB}(\mathbf{q})=&1-\frac{\rho}{\rho_{A}}\delta_{AB} \label{Sq_ld}\\
&+
 \frac{1}{x_Ax_B}F_{A}(-\mathbf{q})F_{B}(\mathbf{q})\frac{(2\pi)^3}
 {\rho V_{\mathrm{WS}}^2}\sum_{\mathbf{G}}\delta^{(3)}
 (\mathbf{q}-\mathbf{G}),\nonumber
\end{align}
where $\mathbf{G}$ is the reciprocal lattice vector of each structure,
$x_A$ ($x_B$) is the fraction of particles of species $A$ ($B$) in the
system, $V_{\mathrm{WS}}$ is the volume of the WS cell, $F_A$ ($F_B$)
is the form factor of a single nucleus for species $A$ ($B$) defined as
\begin{eqnarray}
 F_A(\mathbf{q})\equiv \int_{V_{\mathrm{n}}}
 d^3r\ \rho_{A}^{(N)}(\mathbf{r})\cdot e^{i\mathbf{q}\cdot
 \mathbf{r}}.\label{Fq def}
\end{eqnarray}
Here $\rho_A^{(N)}$ is the difference of densities of species $A$ 
inside and outside nuclei, and the integration $\int_{V_{\mathrm{n}}}$ 
is over the volume $V_{\mathrm{n}}$ of one nucleus.
In the following, we use the above expressions to calculate 
$\langle \overline{S}(E_{\nu})\rangle$
for models given by ourselves \cite{watanabe_liquid,watanabe_liquid_err} and
by Lattimer and Swesty \cite{lseos} in which lattice structure is
treated within the WS approximation.
In the present calculation, however, we take account of the lattice structure
explicitly: we assume a BCC lattice of spherical nuclei and bubbles,
a triangular lattice of rod-like nuclei and bubbles, and a one-dimensional
lattice of slab-like nuclei. 
These assumptions are justified by the fact that
the Coulomb energy for the assumed lattice structure is well reproduced
by the WS approximation \cite{oyamatsu}.
In the following numerical calculations,
the summation about $\mathbf{G}$ in Eq.\ (\ref{Sq_ld})
is done for the six smallest reciprocal
lattice vectors because the contribution of higher orders is negligible
in the present situation, where the nuclear volume fraction is large
(the contributions of the fifth and sixth orders are already $10^{-3}$ 
smaller than that of the first order).

\begin{figure}
\rotatebox{0}{
\resizebox{7cm}{!}
{\includegraphics{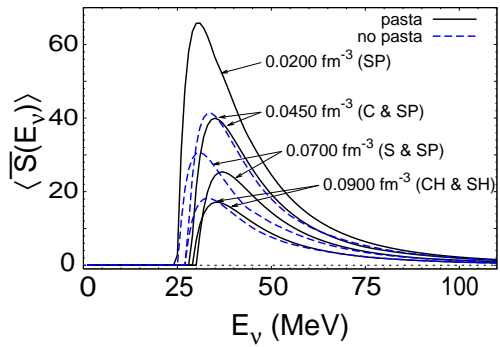}}}
\caption{(Color online) $\langle \overline{S}(E_{\nu})\rangle$ at $Y_L=0.3$ calculated for
 a liquid drop model based on BBP with a typical parameter set given by 
 Ref.\ \cite{watanabe_liquid}.
 The abbreviations SP, C, S, CH, and SH stand for
 phases with spherical nuclei, cylindrical nuclei, slab-like nuclei,
 cylindrical holes and spherical holes, respectively.
 \label{sq_t_bbp_spb_snm}}
\end{figure}

To see the effects of existence of nuclear pasta clearly,
we compare $\langle \overline{S}(E_{\nu})\rangle$ for two cases:
(1) we take account of the pasta phases, i.e., the phases with
rod-like nuclei, slab-like nuclei, and rod-like bubbles, in addition
to the phases with spherical nuclei and bubbles; 
(2) we only take account of the phases with spherical nuclei and bubbles
(here referred to as the no pasta case).
For case (1), we use nuclear data of supernova matter at zero temperature
and at $Y_L=0.3$ obtained in Refs.\ \cite{watanabe_liquid}
and \cite{watanabe_liquid_err}.
In these references we used a liquid drop model, which is based on
that developed by Baym, Bethe, and Pethick (BBP) \cite{bbp}.
Here we use results for standard values of parameters for the surface tension
and the chemical potential in this model.
Case (2) is newly calculated for the same parameters using the same model
and taking account of only spherical nuclei and bubbles.
In Fig.\ \ref{sq_t_bbp_spb_snm} we show the results of these two cases.
We can see the common tendencies that
$\langle \overline{S}(E_{\nu})\rangle$
is zero below $E_\nu\simeq$ 25 MeV, rapidly increases at 25--30 MeV, has a peak
around 30--40 MeV, falls relatively gently, and approaches 
$x_n+(c_v^{(p)}/c_v^{(n)})^2 x_p$ above 100 MeV \cite{note_ld}.
Peak height monotonically decreases with increasing density.
This is simply because the nucleon density distribution becomes more uniform.
Striking differences between the results for the cases with and without
the pasta phases are as follows.
The peak height decreases and the peak energy increases
when one takes account of the pasta phases.
The former is caused by the disappearance of the coherence along
the axis of cylindrical nuclei and in the plane of slab-like nuclei.
The latter is due to smaller values of the lattice constant
for the phases with rod-like and slab-like nuclei compared to those
for the phases with spherical nuclei and bubbles;
this is a general property observed in a number of previous calculations
\cite{lrp,oyamatsu,watanabe_liquid,watanabe_liquid_err,maru_scr}.

\begin{figure}
\rotatebox{0}{
\resizebox{8cm}{4cm}
{\includegraphics{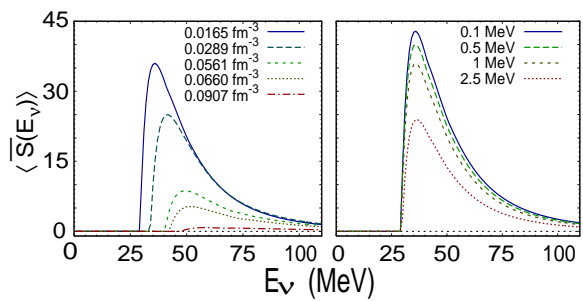}}}
\caption{(Color online) $\langle \overline{S}(E_{\nu})\rangle$ at $x_p=0.3$ calculated
 for EOS by Lattimer and Swesty (LS-EOS) \cite{lseos}.
 In the left panel, the temperature is fixed at $1$ MeV.
 In the right panel, the number density of nucleons is fixed at
 0.0165 $\mathrm{fm^{-3}}$.
 \label{sq_t_ls_eos}}
\end{figure}

In Fig.\ \ref{sq_t_ls_eos}
we show $\langle \overline{S}(E_{\nu})\rangle$ for the equation of state (EOS)
by Lattimer and Swesty (LS-EOS) \cite{lseos} at a proton fraction $x_p=0.3$
to see effects of finite temperatures within the liquid drop approach 
\cite{note_shen}.
The left panel shows the density dependence and the right one
the temperature dependence.
We note that, in this calculation, nuclei are assumed to be spherical
and the other structures including the spherical bubbles 
are not taken into account.
In the present calculation, when we take an average of $S_{AB}(\mathbf{q})$
over the scattering angle, we multiply $S_{AB}(\mathbf{q})$ by the
Debye-Waller factor given by Ref.\ \cite{kaminker} and use the relation
$q^2=2E^2_{\nu}(1-\cos\theta)$.
The right panel shows that $\langle \overline{S}(E_{\nu})\rangle$
falls monotonically with increasing temperature.
This is due to thermal fluctuations of the position of nuclei 
from equilibrium lattice points described by the Debye-Waller factor
and the smoothed nucleon density profile at higher temperatures.
Although these two kinds of finite temperature effects can be (even partially)
incorporated within the liquid drop model such as LS-EOS,
there is another kind of finite temperature effect that is hard to
be described by the liquid drop model: disorder.
Figure \ref{sq_t_ls_eos} exhibits a steep decline of the peak
on the low energy side. Disorder will
broaden the peak of $\langle \overline{S}(E_{\nu})\rangle$ and
will produce a spectral tail on the low energy side since 
the disorder of lattice contributes to $\langle \overline{S}(E_{\nu})\rangle$
at wavelengths necessarily longer than the lattice constant.
The QMD can describe 
the finite temperature effects including disorder 
beyond the limit of liquid drop models.
Next, we show $\langle \overline{S}(E_{\nu})\rangle$ calculated for
QMD simulations as more realistic results.

Here we briefly summarize QMD models and computation techniques.
We use nuclear forces given by Maruyama {\it et al.} \cite{maruyama}
(model 1) and by Chikazumi {\it et al.} \cite{chikazumi} and Kido
{\it et al.} \cite{kido} (model 2)
with medium EOS parameter sets \cite{note_error}.
Model 1 is used in our previous works
\cite{qmd_r,qmd,qmd_hot,qmd_transition}. 
Model 2 is newly adopted for studying the dependence of the phase diagram
on nuclear forces within the framework of QMD \cite{qmd_modeldepend}.
Parameter sets for both the models are determined by
fitting the saturation properties of nuclear matter, 
the properties of finite nuclei in the ground state, 
and the energy dependence of the optical potential.
In the Hamiltonian of model 2,
there is an extra term depending on the density gradient, which
is not included in that of model 1.
Procedures to obtain the equilibrium states of hot nuclear matter
by QMD are explained in Ref.\ \cite{qmd_hot} for model 1 and
in Ref.\ \cite{qmd_modeldepend} for model 2.

We calculate the angle-averaged radial distribution function $g_{AB}(r)$
for snapshots of QMD simulations with 16384 (or 2048) nucleons at a
proton fraction $x_p=0.3$ \cite{note_size}.
In calculating $g_{AB}(r)$,
we place $5^3$ (or $10^3$) resulting simulation boxes, each containing 16384
(or 2048) nucleons, to enlarge the cutoff radius of $g_{AB}(r)$;
we take the cutoff radius to be greater than $100$ fm.
Then we obtain $S_{AB}(q)$ by calculating the Fourier transform
of $g_{AB}(r)$ multiplied by a Hanning window \cite{recipe},
which removes ripples caused by Fourier transforming $g_{AB}(r)$ 
with a discrete change at the finite cut-off radius.

\begin{figure}
\rotatebox{0}{
\resizebox{8cm}{!}
{\includegraphics{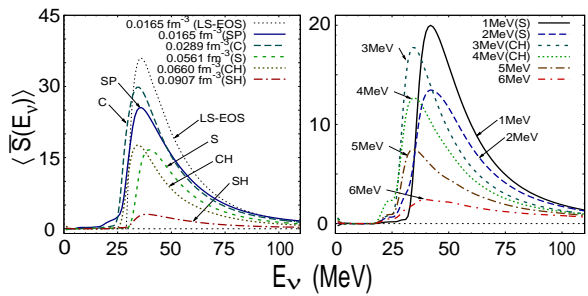}}}
\caption{(Color online) $\langle \overline{S}(E_{\nu})\rangle$ at $x_p=0.3$ calculated
 for QMD model 1 (left panel) and model 2 (right panel).
 Abbreviations are the same as in Fig.\ \ref{sq_t_bbp_spb_snm}.
 The left panel shows $\langle \overline{S}(E_{\nu})\rangle$
 for each pasta phase at $T=1$ MeV.
 For comparison we also show $\langle \overline{S}(E_{\nu})\rangle$
 at 0.0165 $\mathrm{fm}^{-3}$ and $T=1$ MeV for LS-EOS.
 The right panel shows $\langle \overline{S}(E_{\nu})\rangle$ for
 various temperatures at a fixed density 0.0660 $\mathrm{fm}^{-3}$.
 \label{sq_t_qmd}}
\end{figure}

In Fig.\ \ref{sq_t_qmd}
we show $\langle \overline{S}(E_{\nu})\rangle$ 
at a fixed temperature ($T=1$ MeV) for QMD model 1 (left panel) and
at a fixed density (0.0660 $\mathrm{fm}^{-3}$) 
for QMD model 2 (right panel), respectively.  
From these figures, we observe two important characteristics of 
$\langle \overline{S}(E_{\nu})\rangle$
calculated for the results of the QMD simulations.
One is that the peak is lower and broader than the results for 
the liquid drop models shown before
(see the lines for SP and LS-EOS in the left panel of Fig.\ \ref{sq_t_qmd}).
This is due to lattice disorder and irregularities of nuclei,
which can be incorporated by QMD, but not by the liquid drop models.
The former (latter) enhances the cross section of low (high) 
energy neutrinos, while they reduce that of neutrinos around peak energy.
It is noticeable that, due to the former effect, 
$\langle \overline{S}(E_{\nu})\rangle$ calculated for the QMD results
has nonzero values in the low energy region of $E_\nu \alt 25$ MeV
in some cases while $\langle \overline{S}(E_{\nu})\rangle$ shown in
Figs.\ \ref{sq_t_bbp_spb_snm} and \ref{sq_t_ls_eos} vanishes in this region.
The other important characteristic
is that, as shown in the left panel of Fig.\ \ref{sq_t_qmd},
the peak height does not necessarily decrease with
increasing density unlike the results for the liquid drop models
shown in Figs.\ \ref{sq_t_bbp_spb_snm} and \ref{sq_t_ls_eos}.
This nonmonotonic behavior is observed between the phases with 
spherical nuclei and rod-like ones, and also between the phases with 
slab-like nuclei and rod-like bubbles.
Let us first consider the latter case.
In the results of the QMD simulations, slab-like nuclei form a layered lattice
and rod-like bubbles form a triangular lattice.
The ratio of the number of the smallest reciprocal lattice vectors
of a layered lattice to that of a triangular one is 1:3.
This larger number of the lowest order of the reciprocal lattice vectors
for the phase with rod-like bubbles might explain the larger peak height
compared to the phase with slab-like nuclei \cite{note_size}.
For the same reason, it is possible that
$\langle \overline{S}(E_{\nu})\rangle$ increases with a transition 
from the phase with slab-like nuclei to that with rod-like bubbles 
caused by an increase of temperature
as suggested in the right panel of Fig.\ \ref{sq_t_qmd}. 
Here nuclei adopt a slab-like shape at $1$--$2$ MeV
and nuclear structure transforms to rod-like bubbles at $3$--$4$ MeV,
above which nuclei deform and nuclear surface cannot be identified.
On the other hand, the reversal of the peak height between the phases
with spherical nuclei and rod-like ones 
would be explained by
the position of spherical nuclei in the former being easy to fluctuate
compared to that of rod-like nuclei in the latter, 
where the nuclear matter region is connected in one direction 
and this acts to suppress the fluctuation in the transverse direction.

\begin{table}
\caption{The mass of the region where pasta phases would appear
($M_{\mathrm{pasta}}$) and that where nuclei exist ($M_{\mathrm{nuc}}$)
just before the bounce in the typical supernova simulations
for $T/|W|=0$ and 0.01. Radius means the distance of possible regions for
pasta phases from the center of the core. Mass is measured by the solar
mass $M_{\odot}$. Ratio is defined as
$M_{\mathrm{pasta}}/M_{\mathrm{nuc}}$.}
\label{tab:mass_pasta}
\begin{tabular}{c|cccc}
 $T/|W|$& Radius (cm) & $M_{\mathrm{pasta}}(M_{\odot})$ &
 $M_{\mathrm{nuc}} (M_{\odot})$& Ratio\\ \hline
 0 & 1.1--1.3 $\times 10^6$  &0.13
 & 0.92 &0.14\\
 0.01 & 0.8--1.6$\times 10^6$  & 0.30&1.29&0.23\\ \hline
\end{tabular}
\end{table}

Finally we estimate the possible mass of the pasta phases in supernova
cores to examine their importance on the core-collapse supernova explosions.
We have performed two-dimensional axisymmetric simulations of
rotational core collapse supernovae.
The initial condition is taken from the profile of the 15 $M_{\odot}$
nonrotating progenitor with solar metalicity at zero-age main sequence
(ZAMS) by Woosley {\it et al.} \cite{whw_rev}.
The mass of the core is 1.58 $M_{\odot}$.
Rotation is then added in the same way as in Ref.\ \cite{takiwaki}.
Here we set $T/|W|=0$ and 0.01, where
$T/|W|$ is a ratio of the rotational kinetic energy $T$
to the gravitaional binding energy $W$ of the core.
These values of $T/|W|$ are typical and the latter one is also
suggested by simulations of the evolution of massive stars
with a typical rotational velocity on the equator at ZAMS,
$\sim 200 \mathrm{km}/\mathrm{s}$ \cite{whw_rev,hlw_evol}.
The magnetic field is set to be zero.
We use an EOS by Shen {\it et al.} \cite{sheneos}.
Other basic equations and input physics are shown in Refs.\ 
\cite{takiwaki} and \cite{kotake}.

In Table \ref{tab:mass_pasta}, we show the mass $M_{\mathrm{pasta}}$
of the pasta phases just before the bounce.
The region where the matter consists of the pasta phases is 
defined by the following two conditions: 1) nuclei exist and
2) the volume fraction of nuclei is larger than $1/8$,
which is a condition for the fission instability of
spherical nuclei giving a reasonable estimate of the density
at which matter starts to consist of nonspherical nuclei
\cite{pr_review}.
Note that the EOS used in our simulations
only incorporates the phases with spherical nuclei
and uniform nuclear matter and it does not take account of the phase with
spherical bubbles and any other pasta phases.
Therefore values
of $M_{\mathrm{pasta}}$ in Table \ref{tab:mass_pasta} are lower
limits in our simulations: some region of uniform nuclear matter
close to a boundary with the phase with nuclei in our simulations
corresponds to the pasta phases.
We also show the mass $M_{\rm nuc}$ of the region where nuclei exist
in the core because the pasta phases modify the cross section of
elastic neutrino scattering, which is dominated by the contribution of nuclei.
It is notable that about 10--20 \% of spherical nuclei are
replaced by the pasta nuclei in both the rotating and no-rotating cases.
In addition, the moderate rotation in the core extend the region of pasta
phases by centrifugal force \cite{note_rot}.

In summary we have shown how pasta phases affect energy-dependent 
cross sections for coherent scattering of neutrinos in collapsing cores.
At zero temperature, existence of the pasta phases instead of
the phases with spherical neuclei and bubbles increases the energy
of the peak of the cross section at $E_\nu \simeq 30$--$40$ MeV and 
decreases the peak height.
They lead to a reduction in the opacity of the core 
for low energy neutrinos of $E_\nu \alt 30$ MeV.
At non-zero temperatures, the peak height further decreases
but increases the cross section of neutrinos at energy lower and higher
than that of the peak.
The pasta phases, which amount to 10--20 \% of the mass of supernova cores,
would modify the lepton fraction of the core.
The next step is to construct a systematic numerical table 
or a useful fitting formula of the cross section for 
elastic neutrino scattering in the pasta phases and perform
hydrodynamic core collapse simulations, which incorporate 
the effects of the pasta phases on the neutrino cross section
studied in the present article.

\vspace{3mm}
We are grateful to T. Maruyama, S. Chikazumi, K. Iida, and K. Oyamatsu 
for helpful discussions.
We also thanks Chris Pethick for valuable comments.
This work was supported in part
by JSPS,
by the Junior Research Associate Program in RIKEN
through Research Grant No. J130026,
by Grants-in-Aid for Scientific Research provided by the
Ministry of Education, Culture, Sports,
Science and Technology through Research Grants No. 14-7939
and No. 14079202,
and by the Nishina Memorial Foundation.
Parts of the simulations were performed by the RIKEN 
Super Combined Cluster System.

\bibliography{neu-pasta_let.bib}

\begin{thebibliography}{42}
\bibitem[{not({\natexlab{a}})}]{note_elastic}
\bibinfo{note}{Here we are interested in coherent scattering of neutrinos from
  heavy nuclei at subnuclear densities and thus a typical value of the momentum
  transfer $\hbar q$ is $q r_{N}\alt 1$ and that of the energy transfer
  $\hbar\omega$ may be given by $\sim \hbar q v_T \sim \hbar q (T/mA)^{1/2}$,
  where $v_T$ is the thermal velocity of heavy nuclei, $T$ is the temperature,
  and $m$ is the mass of nucleons. Therefore the energy transfer relative to
  the thermal energy is $\delta\equiv \hbar\omega/T\alt \hbar r_{N}^{-1}
  (1/mAT)^{1/2} \sim 0.6 A^{-1/2}$ for $r_{N}=10$ fm and $T=1$ MeV. Provided
  $\delta$ is much smaller than unity, the assumption of elastic scattering is
  valid and effects of degeneracy of neutrinos can be also neglected
  \cite{lamb}. However, at higher densities, where nucleons are strongly
  degenerate and the Fermi velocity of nucleons $v_F\sim c$, a typical energy
  transfer is $\sim \hbar q v_F$ and $\delta$ is no longer small for the above
  values of $q$ and $T$. Thus, in this case, the elastic assumption is not
  accurate \cite{reddy,burrows} and one should also take account of effects of
  degeneracy of neutrinos.}
\bibitem[{not({\natexlab{b}})}]{note_ld}
\bibinfo{note}{Because here we employ a liquid drop model in which the density
  profile of nuclei is approximated by a rectangular form, the resulting
  $\langle \overline{S}(E_{\nu})\rangle$ is reliable only if the wavelength of
  neutrinos $\lambda_\nu$ is comparable to or larger than the typical size of
  the rectangular profile, $\lambda_\nu\agt r_{N}$ ($r_{N}$ is the half width
  of nuclei). For, e.g., $r_{N}=10$ fm, this condition reads $E_\nu \alt 120$
  MeV.}
\bibitem[{not({\natexlab{c}})}]{note_shen}
\bibinfo{note}{We have also calculated $\langle \overline{S}(E_{\nu})\rangle$
  for another macroscopic model by Shen {\it et al.} \cite{sheneos}. The
  following discussion also holds for this model.}
\bibitem[{not({\natexlab{d}})}]{note_error}
\bibinfo{note}{We remark that the expression for the Hamiltonian given in Ref.\
  \cite{chikazumi} is erroneous. The quantity $L$ in the density dependent term
  [the second term in Eq.\ (2)] of that paper should be replaced by $\tilde{L}$
  defined in the same way as in Ref.\ \cite{maruyama} (private communication
  with T. Maruyama).}
\bibitem[{not({\natexlab{f}})}]{note_rot}
\bibinfo{note}{Rapid rotation in the core drives the bounce by the strong
  centrifugal force, not by the nuclear repulsive force. In this case the
  central density just before the bounce is too small for the pasta phases to
  appear.}
\end{thebibliography}

\end{document}